\documentclass{elsart}
\usepackage{epsfig}
\usepackage{psfrag}
\usepackage{graphics}
\usepackage{graphicx}
\usepackage{amssymb}
\usepackage{dcolumn}

\newcommand{\difrac}[2]{\frac{\displaystyle #1}{\displaystyle #2}}
\newcolumntype{d}[1]{D{.}{.}{#1}}

\begin{document}

\begin{frontmatter}
\begin{center}

\title{Time and space reconstruction in optical, non-imaging, scintillator-based particle detectors}
\author[pu-phys]{C. Galbiati},
\author[pu-phys]{K. McCarty\corauthref{cor1}}
\corauth[cor1]{Corresponding author. Tel: +1 (609) 258-4330}
\ead{kmccarty@princeton.edu}
\address[pu-phys]{Physics Department, Princeton University, Princeton, NJ 08544, USA}

\begin{abstract}
A new generation of ultra-low-background scintillator-based detectors aims to
study solar neutrinos and search for dark matter and new physics
beyond the Standard Model.  These optical, non-imaging detectors
generally contain a ``fiducial volume'' from which data are accepted, and
an ``active buffer region'' where there are higher levels of radioactive
contaminants.  Events are observed in real time.   To
distinguish between events occurring in the two regions, it is imperative that
event position reconstruction be well-understood.  The object of this paper is
the study of the reconstruction, in time and space, of scintillation events in
detectors of large dimensions.  A general, likelihood-based method of position
reconstruction for this class of detectors is presented.  The potential spatial
resolution of the method is then evaluated.  It is shown that for a spherical
detector with a large number $N$ of photosensitive elements that detect
photons, the expected
spatial resolution at the center of the detector is given by $\delta a \approx
(c \sigma / n) \sqrt{3/N}$, where $\sigma$ is the width of the scintillator
time response function and $n$ is the index of refraction in the medium.
However, if light in the detector has a scattering mean free path
much less than the detector radius $R$, the resolution
instead becomes $(R/2)\sqrt{3/N}$.
Finally, a formalism is introduced to deal with the common case in which
only the arrival time of the first photon to arrive at each photosensitive
element can be measured.
\end{abstract}


\end{center}
\end{frontmatter}

\section{Introduction}

Optical, non imaging detectors are widely used for the detection of weakly
interacting particles.  At present the main focus of observation is on
neutrinos and antineutrinos from various sources, but there are also plans to
construct large optical detectors to search for as yet undiscovered particles
such as WIMPs.  The detection mechanism is based on the collection of visible
or ultraviolet photons.  These are emitted as \v{C}erenkov radiation ({\it
e.g.}, as in Kamiokande~\cite{super-k} and SNO~\cite{sno}) or as
scintillation photons.  We will focus our attention in this paper on
scintillator-based, unsegmented detectors.

\subsection{A Brief History of Scintillation Detectors}

The history of scintillator-based detectors is heavily intertwined with that of
neutrino physics.  The first neutrino detector ever built, that of Cowan and
Reines in 1953, was a 10.7\,ft$^3$ cylinder filled with a cadmium-doped organic
scintillator and wavelength shifter, which detected
reactor-generated $\bar{\nu_e}$'s by observing the coincidence of $e^+$
annihilation and neutron capture following the inverse beta decay reaction
$\bar{\nu_e}(p, n)e^+$~\cite{cr-scint,cr-discovery,cr-confirmation}.
However, the first large-scale unsegmented liquid scintillator detector was not
built until about 1980.  The 100~ton neutrino detector at Artemovsk, Ukraine, a
cylindrical 5.6\,m $\times$ 5.6\,m tank filled with a saturated hydrocarbon
scintillator and fluor, was a direct descendant of Reines and Cowan's original
design.  Indeed, it was designed to detect antineutrinos using the same
reactions~\cite{beresnev}.  It was buried in a salt mine, 600 meters water
equivalent (m.w.e.) underground, and was later used to study the
interactions of cosmic ray muons with scintillator~\cite{enikeev}.

The 1995 Counting Test Facility (CTF) prototype of the Borexino experiment
further developed the architecture of scintillator-based detectors~\cite{ctf}.
This 4~ton detector was intended primarily as a test bed for technologies of
the full-scale Borexino detector, not as a neutrino detector in its own right.
Nevertheless, it set a record for the lowest detector background achieved
at the time, of 0.03\,counts/(kg\,keV\,yr), in the window 250\,keV to
2.5\,MeV~\cite{ctf-results}.  It has as a result produced new upper bounds on
various exotic processes~\cite{ctf-exotic}.  Unlike previous scintillation
detectors, it is spherical in design, in order to keep as much
scintillator away from the surface as possible.  Liquid scintillator (both
pseudocumene and phenylxylylethane, at different times, again with added
fluors) is contained in a thin spherical nylon balloon, surrounded by 100
inward-facing photomultiplier tubes.  This setup is contained in 1000~tons of
ultrapure water in a cylindrical tank.  The entire detector is 3400~m.w.e.
underground in the Gran Sasso National Laboratory, in central Italy.
The CTF first established the feasibility of a scintillator-based {\it solar}
neutrino detector with a detailed study of the radioactive contaminants
internal to the scintillator.  It was also the first scintillation detector to
introduce an inactive buffer (water) between the active volume of scintillator
and the photomultiplier tubes.  As well, it has the capability of position
reconstruction for point-like events.

The CHOOZ detector~\cite{chooz}, built to study oscillations in reactor
antineutrinos from a nuclear power plant by the same name in northern France,
took data in 1997-98.  Its layered design incorporated key features of the CTF
and Borexino designs, as well as those of other neutrino detectors such as the
\v{C}erenkov detector SNO~\cite{sno}, and the hybrid \v{C}erenkov/scintillation
light detector LSND~\cite{lsnd}.  (The design of the larger hybrid detector
MiniBooNE, built in 1999 in order to confirm or refute results from LSND by
observing 0.5-1\,GeV muon neutrinos produced at the FNAL accelerator,
was based upon the same principles~\cite{miniboone}.)
The interior of the CHOOZ detector featured a
central 5~ton Gd-doped target mass inside a clear roughly egg-shaped Plexiglas
container, surrounded by an undoped 17~ton inactive buffer region contained in
an oblong ``geode,'' and an outer undoped 90~ton volume with its own set of
PMTs, used for vetoing muons from cosmic rays.  The detector was placed
at a depth of 300~m.w.e.

The current generation of unsegmented detectors based on organic liquid scintillators -
KamLAND~\cite{kamland}, taking data since 2002, and Borexino~\cite{bx}, soon to
begin operations - retain this sort of layered design, both using the spherical
shape of the CTF.  Unlike the detectors described already, Borexino will
observe scintillation light due directly to neutrino scattering from electrons,
and can therefore potentially detect neutrinos with much lower energies (the
threshold $\nu$ energy for the inverse $\beta$ decay is 1.8\,MeV).  KamLAND has
observed disappearance of $\bar{\nu_e}$ from reactors using the inverse
$\beta$ decay signature~\cite{kamland-results}, but it is also intended to
observe solar neutrinos directly via
$\nu$-$e$ scattering in the future.  The current KamLAND background in the
region below 2\,MeV must be drastically reduced for that goal to be
achieved~\cite{kamland-background}.  These detectors are
situated much deeper underground
(Borexino: 3400 m.w.e; KamLAND: 2700 m.w.e.), for further reduction of the
residual muon flux and the production of short-lived cosmogenic isotopes.

Two new experiments with targets of liquified noble gas, also aiming at low
energy solar neutrino detection via the detection of scintillation light,
are currently under development: CLEAN~\cite{clean} and XMASS~\cite{xmass}.
In the case of CLEAN, wavelength-shifter coated windows are offset from the
PMTs by a 5-10\,cm gap which is a thin inactive buffer region.
For these detectors, reliable determination
of the positions of events is even more important, due to the need of rejecting
the higher background rate coming from scintillation events produced in
proximity of the PMTs and container vessel.  An additional complication arises because the mean
scattering length of scintillation photons (produced in the ultraviolet range
of the spectrum for noble gases) is much less than the radius of the detector;
scintillation photons propagate from the event origin to the PMTs in a
diffusive mode.  Therefore the times of arrival of detected photons provide
less information than in detectors using organic scintillator; these 
noble gas detectors will rely heavily upon the
spatial pattern of PMT hits to reconstruct the positions of events.

\subsection{The Necessity of Spatial Event Reconstruction}

Due to the extremely low interaction rates of neutrinos and their antiparticles
(to say nothing of WIMPs and so forth), it is necessary for a detector to
contain a large mass of scintillator with very low levels of internal
radioactive contamination~\cite{bx}.  Ultra-pure materials are also used to
screen radioactivity from materials surrounding the detector~\cite{bx,bx-rad}.
Unfortunately, the photosensitive elements used to detect scintillation light
are notorious for being among the main sources of radioactivity in an
ultra-low-background detector.

It is therefore desirable to insert, between the photosensitive elements and
the scintillator, one or more layers of buffer material to suppress radioactive
background.  Often the buffers are inactive, {\it i.e.}, not scintillating.  An
inactive buffer offers the advantage of minimizing the total trigger rate
caused by the abundant radioactive decays generally produced within the
photosensitive elements~\cite{bx}.  On the other hand, if the compositions of
the scintillator and inactive buffer are different, a scintillator containment
system is required to physically separate them~\cite{bx}.  The
containment system, being in direct contact with the scintillator,
must satisfy extremely stringent requirements in terms of intrinsic
radiopurity.

For additional background prevention, the outer region of the scintillator
volume can be used as an active buffer.  This allows any residual radioactivity
coming from the containment system, or passing through it, to be monitored and
suppressed.  A ``fiducial volume'' is commonly defined as a region at
the center of the active volume of the detector in which radioactive background
is expected to be at a minimum.  The discrimination between
events belonging to the fiducial and to the non-fiducial regions is performed
by means of software implementation (reconstruction code) of an algorithm
(reconstruction algorithm), which assigns to each single event a reconstructed
position, either inside or outside the fiducial volume. The algorithm also
provides a means of comparing the position of different events and is an
important tool for the identification of several background sources.  The
designs of some planned detectors incorporate only a thin inactive buffer
region or none at all, and in these cases, correct assignment of
an event as belonging to the fiducial volume or the buffer region is even more
important.  The resolution of detector reconstruction codes are generally
studied with Monte Carlo methods.  Event simulations allow close reproductions
of the performance of these codes on real events.  Typically, however, the
reconstruction codes are fine tuned by calibrating the detector with the use of
localized sources of radioactivity or light.

What seems lacking from the available literature is a comprehensive discussion
of how the resolutions of detector reconstruction codes are related to some
basic properties of the detector: the linear dimension, the time dispersion of
the photon emission, the scintillator index of refraction, possible processes
of absorption and re-emission and of scattering of the scintillator light, etc.
In this paper we present an analytic study of the resolution for reconstruction
in time and space of scintillation events.  The study is restricted, for
simplicity, to the case of events at the center of the detector, simple enough
to be treated, within certain approximation, analytically.  Calibrations of
experiments~\cite{ctf-results} and full Monte Carlo studies of the performance
of proposed experiments~\cite{ctf-light} show anyhow that the resolution of the
reconstruction codes depends only in a mild way upon the location of the
scintillation event.

This study also assumes that the optical properties of the media are uniform
throughout the detector, and that the indices of refraction of all materials
between the active scintillator and the photodetectors are approximately the
same.

\section{Likelihood Function Derivation}

The likelihood function is a standard statistical tool used for finding parameters of a physical model.  Suppose that a set of $N$ observations is composed of the independent values $\{t_i\}$ and dependent values $\{s_i\}$ ($i = 1, ..., N$).  For instance, $\{t_i\}$ could be a list of times at which a radioactive sample is observed, and $\{s_i\}$ a list of observed activities at each time.  We wish to model the data using some function $f(s)$ with $n$ free parameters $\vec{a}$.  In the example, the function would be a decaying exponential, and the parameters would be the initial activity and the half-life.  By definition, the likelihood function over the parameters is a probability distribution of obtaining 
the observed data given a specific set of parameters:

\begin{equation}
\mathcal{L}(\vec{a_0};\, \{(t_i, s_i)\})
\; = \; \mathrm{P}(\{(t_i, s_i)\} \; \mathrm{are\; observed}
	\;|\; \vec{a} \;=\; \vec{a_0}).
	\label{e:base-likelihood}
\end{equation}

The difficult task is to calculate this probability based on the assumption
that the data are correctly described by the model function $f(s)$.  Once
this has been done, in order to calculate the most probable value of the
parameters of the model, one simply finds the maximum of the likelihood
function (or, as is usually computationally easier, the minimum of $-\log
\mathcal{L}$) in the $n$-dimensional space defined by the free parameters
$\vec{a}$.

In the case of a scintillator-based detector, the parameters of interest are the position and time of an event in the detector, $\vec{a} = (\vec{x_0}, t_0)$.  The observed data are the positions $\{\vec{x_i}\}$ of the
photosensitive elements, usually PMTs (independent values), and the times $\{t_i\}$ at which each element is hit by a photon (dependent values); $i$ ranges from 1 to $N$,
with $N$ being the number of detected photons.  For now we assume that at most one photon is detected by each PMT, so all the $\vec{x_i}$'s are distinct, and $N$ is also the number of PMTs that detect a photon.
For conciseness, define the following possible events:
\begin{itemize}
\item{A : detector event occurs at $(\vec{x_0}, t_0)$}
\item{B : detector hit pattern is $\{(\vec{x_i}, t_i)\}$.}
\end{itemize}
Then, Equation~(\ref{e:base-likelihood}) becomes
\begin{equation}
\mathcal{L}(\vec{x_0}, t_0; \{(\vec{x_i}, t_i)\}) \; \equiv \;
\mathrm{P}(\mathrm{B} | \mathrm{A}).
	\label{e:bayes}
\end{equation}

\subsection{Factoring the Detector Likelihood Function}

Let us assume that the times at which photons are emitted by the scintillator
are uncorrelated.  Then the likelihood function will have one independent
factor for the piece of data provided by each PMT\footnote{
Strictly speaking, this is not exactly true; specifying that $N$
PMTs detected photons causes the PMT hit data to be correlated. For a
reasonably large number of hit PMTs, though, the difference should be
negligible.  It would be interesting to compare results derived from
the often-used Poisson and multinomial probabilistic models to the model
put forth here}.  Let the total number of working PMTs be $T$, so
that $N$ PMTs (labeled $1, \ldots, N$) have detected a
photon, and $T - N$ PMTs (labeled $N + 1, \ldots, T$) have not.  If we
further define
\begin{itemize}
\item{C$_i$ : PMT $i$ is hit}
\item{D$_i$ : PMT $i$ detects a photon}
\item{E$_i$ : PMT $i$ detects a photon at time $t_i$,}
\end{itemize}
then
\begin{eqnarray}
\mathrm{P}(\mathrm{B} | \mathrm{A})\; & = & \;
\prod_{i = 1}^N
\mathrm{P}(\mathrm{E}_i | \mathrm{A}, \, \mathrm{C}_i, \, \mathrm{D}_i)\;
\mathrm{P}(\mathrm{D}_i | \mathrm{A}, \, \mathrm{C}_i)\;
\mathrm{P}(\mathrm{C}_i | \mathrm{A}) \nonumber \\
& & \times\; \prod_{j = N+1}^T \left[
\mathrm{P}(\neg \mathrm{D}_j | \mathrm{A}, \, \mathrm{C}_j)\;
\mathrm{P}(\mathrm{C}_j | \mathrm{A})\; + \;
\mathrm{P}(\neg \mathrm{C}_j | \mathrm{A}) \right]
\label{e:likelihood-factors}
\end{eqnarray}
(where $\neg$ is the logical negation symbol).
Of course, $\mathrm{P}(\mathrm{D}_i | \mathrm{A}, \, \mathrm{C}_i)$ is just
the quantum efficiency $q_i$ of PMT $i$, which is, to a first
approximation, independent of the original event position.

Now define a ``per-PMT'' likelihood function $\mathcal{L}_i$.
\begin{equation}
\mathcal{L}_i(\vec{x_0}, t_0; \vec{x_i}, t_i) \; = \; \left\{
\begin{array}{ll}
q_i\, \mathrm{P}(\mathrm{E}_i | \mathrm{A}, \, \mathrm{C}_i, \, \mathrm{D}_i)\;
\mathrm{P}(\mathrm{C}_i | \mathrm{A}), & i \le N
\\
(1 - q_i) \,
\mathrm{P}(\mathrm{C}_i | \mathrm{A})\; + \;
\mathrm{P}(\neg \mathrm{C}_i | \mathrm{A}), & N < i \le T
\end{array}
\right.
\label{e:per-pmt}
\end{equation}

The total likelihood function is then the product of all per-PMT likelihood
functions.  Notice that the per-PMT
likelihood function of a supposedly dead PMT ($q_i = 0$) that does not detect a
photon reduces to 1, so does not influence the total likelihood function, just
as expected.

\subsection{Scintillator Dispersion Time at the Emission Point}

The first factor in the expression for the likelihood function of a PMT that
detects a photon is based solely on timing information of a photon emitted by
the scintillator.
Scintillation photons are emitted as a consequence of the ionization of the scintillator due to interacting particles or radioactive decays. The typical dispersion in the time of emission of organic liquid scintillators is on the order of a few nanoseconds, with a slower component that can reach hundreds of nanoseconds.  The emission of photons is uniform over the solid angle.  In this discussion we assume that the time of emission of each photon, relative to the time of the event causing scintillation, is an independent random variable $\tau_e$.

\begin{figure}[t!]
\begin{center}
 \psfrag{nhat}{$\hat{n}$}
 \psfrag{tau}{$\tau_e = t - t_0 - \tau_f^i$}
 \psfrag{pdftau}{$p(\tau_e)$}
 \psfrag{event}{$(\vec{x_0}, t_0)$}
 \psfrag{detector}{$(\vec{x_i}, t_i)$}
 \psfrag{da}{$\mathrm{d}A_i$}
 \psfrag{domega}{$\mathrm{d}\Omega_i$}
 \psfrag{psi}{$\psi_i$}
 \epsfig{file=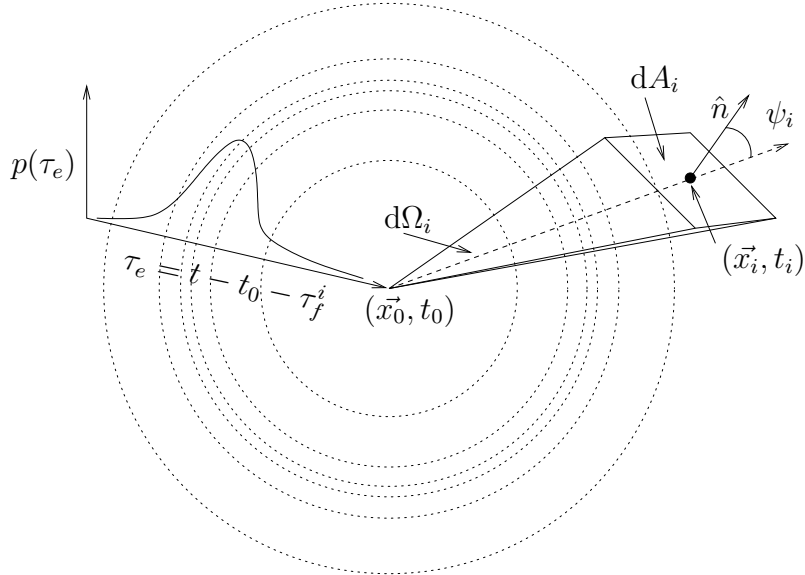,height=3in}
\end{center}
\caption{Geometry of the likelihood function derivation.  The concentric
dotted lines, and the graph on the left, represent the probability function
(an expanding spherical wave) of the emission time of a scintillation photon.  The rectangle labeled d$A_i$ represents a PMT of infinitesimal size with normal vector $\hat{n}$, subtending a solid angle d$\Omega_i$ as seen from the position of the detector event.  The PMT is tilted away from the direction of the event by an angle $\psi_i$.  Note that we have
not yet made any assumptions about the geometry of the detector.}
\label{f:scintpdf}
\end{figure}

Suppose the distribution of the random variable $\tau_e$ is given by some
scintillator response function $p(\tau_e)$.  Referring to the left half of
Figure~\ref{f:scintpdf}, one sees that at a specific time $t$, this function
may also be regarded as an outgoing spherical photon probability wave,
integrated over the solid angle 4$\pi$.  In fact, the most important factor in
Equation~(\ref{e:likelihood-factors}), the probability $\mathrm{P}(\mathrm{E}_i
| \mathrm{A}, \, \mathrm{C}_i, \, \mathrm{D}_i)$, is equal to it.  Let
$\tau_f^i$ be the time of flight from the origin $\vec{x_0}$ of the photon to
the position $\vec{x_i}$ of the $i^{th}$ PMT.  Then, with $n$ being the
scintillator index of refraction, we have:
\begin{eqnarray}
\tau_f^i & = & \difrac{\left|\vec{x_i}-\vec{x_0}\right|n}{c} \\
t_i & = & \tau_e + \tau_f^i + t_0.
\end{eqnarray}
As a result,
\begin{equation}
\mathcal{L}_i(\vec{x_0}, t_0; \vec{x_i}, t_i) \; \propto \;
	p(t_i - t_0 - \tau_f^i).
\end{equation}

Of course, factors other than the dispersion time of the scintillator may also
affect the probability distribution function of the recorded arrival times
of photons at PMTs.  The most important other effects are usually the
effects of scattering in the scintillator and the
finite time resolution of the PMTs themselves.  The latter may in general be
incorporated into the distribution $p(\tau_e)$ by convolution with the
scintillator dispersion function.  The former requires a bit more care
because scattering effects depend in general upon the light path length
from the event to the PMT; an exact treatment is beyond the scope of this
paper.

\subsection{Photon Attenuation}

As photons travel away from their origin, they are attenuated by the
familiar inverse square law.  This implies a formula for the probability
$\mathrm{P}(\mathrm{C}_i | \mathrm{A})$ that
a given PMT is hit by a scintillation photon.  Suppose a PMT of infinitesimal
area,
at a distance $s_i \equiv \left| \vec{x_i}-\vec{x_0} \right|$ from
the event, subtends a solid angle d$\Omega_i$ as seen from the event location.
Assuming a perfect collection efficiency, it will collect only a fraction
d$\Omega_i / 4\pi$ of all photons emitted.  So if $\Gamma$ photons were
emitted, its probability of being struck by at least one of them is
\begin{equation}
\mathrm{P}(\mathrm{C}_i | \mathrm{A}) \; = \;
1 - \left(1 - \frac{\mathrm{d}\Omega_i}{4\pi}\right)^\Gamma
\; \approx \; \Gamma \frac{\mathrm{d}\Omega_i}{4\pi}.
\end{equation}

If the $i^{th}$ PMT has an area d$A_i$ and is tilted away from
the line of sight by an angle $\psi_i$, as shown on the right half of
Figure~\ref{f:scintpdf}, then
$\mathrm{d}\Omega_i = \cos{\psi_i}\, \mathrm{d}A_i / s_i^2$, so
the resulting factor in the likelihood function is given by
\begin{equation}
\mathcal{L}_i(\vec{x_0}, t_0; \vec{x_i}, t_i) \; \propto \;
	\Gamma \difrac{\mathrm{d}\Omega_i}{4\pi} \; = \;
	\Gamma \difrac{\cos{\psi_i}}{4\pi s_i^2}\, \mathrm{d}A_i.
\end{equation}

As mentioned already, all constant factors in a likelihood function may
be discarded with no effect on the location in parameter space of its
maximum.  (To first order, this includes the quantum efficiency $q_i$ of
each PMT.)  The per-PMT likelihood function for a PMT detecting a photon may
thus be redefined as
\begin{equation}
\mathcal{L}_i(\vec{x_0}, t_0; \vec{x_i}, t_i) = 
	p(t_i - t_0 - \tau_f^i) \, \difrac{\cos{\psi_i}}{s_i^2}.
	\label{e:likelihood}
\end{equation}
Its logarithm is
\begin{equation}
\log \mathcal{L}_i \; = \; \log p(t_i - t_0 - \tau_f^i) \; + \;
        \log \cos \psi_i \;-\; 2\log s_i.
\label{e:log-likelihood}
\end{equation}

\subsection{The PMTs Not Triggered}

For completeness, we now consider the case of a PMT that does not detect
a photon produced by an event in the detector.  Its per-PMT likelihood
function, from Equation~(\ref{e:per-pmt}), is given by
\begin{eqnarray}
\mathcal{L}_i(\vec{x_0}, t_0)\, \mathrm{d}^3 \vec{x}\, \mathrm{d}t \;&=&\;
(1 - q_i) \mathrm{P}(\mathrm{C}_i | \mathrm{A})\; + \;
\mathrm{P}(\neg \mathrm{C}_i | \mathrm{A}) \nonumber \\
&=&\; (1 - q_i) \left[1 - \left(1 -
        \frac{\mathrm{d}\Omega_i}{4\pi}\right)^\Gamma \right]
        \; + \; \left(1 - \frac{\mathrm{d}\Omega_i}{4\pi}\right)^\Gamma
        \nonumber \\
&=&\; 1 - q_i + q_i \left(1 - \frac{\mathrm{d}\Omega_i}{4\pi}\right)^\Gamma
        \nonumber \\
&\approx&\; 1 - q_i \Gamma \frac{\mathrm{d}\Omega_i}{4\pi}.
\end{eqnarray}

The logarithm of this per-PMT likelihood function is $\approx \; -q_i \Gamma
\mathrm{d}\Omega_i / 4\pi$.  This term, containing an infinitesimal, is
negligible in size compared to the terms of Equation~(\ref{e:log-likelihood})
coming from per-PMT likelihood functions for PMTs that have detected a photon.
If PMTs are in fact very small compared to any other relevant dimensions of the
detector, it may therefore be ignored.

\subsection{Specialization to a Spherical Detector}

\begin{figure}[t!]
\begin{center}
	\psfrag{psi}{$\psi_i$}
	\psfrag{theta}{$\theta_i$}
	\psfrag{R}{$R$}
	\psfrag{a}{$a$}
	\psfrag{s}{$s_i$}
	\psfrag{O}{$\mathrm{O}$}
	\psfrag{C}{$\mathrm{C}$}
	\psfrag{event}{$\mathrm{A} = (\vec{x_0}, t_0)$}
	\psfrag{detector}{$\mathrm{B} = (\vec{x_i}, t_i)$}
	\epsfig{file=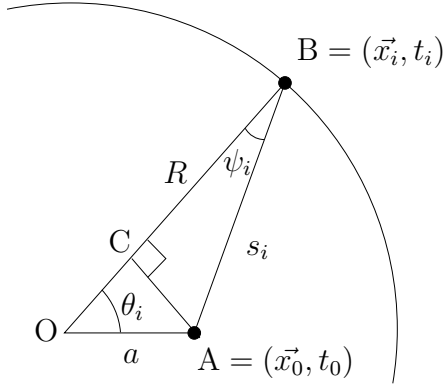,height=2in}
\end{center}
\caption{Geometry of a spherical detector.}
\label{f:circle}
\end{figure}

As written, Equation~(\ref{e:likelihood}) is applicable to any detector with
pointlike PMTs forming the vertices of a convex polyhedron (so that light from an event at any
point inside the detector may reach any one of the PMTs).  Let us specialize to
a spherical detector of radius $R$ centered at the origin, having a uniform
distribution of inward-facing PMTs over the surface.  As above, we call the
distance from an event to the $i^{th}$ PMT $s_i \equiv \left| \vec{x_i} -
\vec{x_0} \right|$.  Let the distance from the center of the detector to the
event be $a \equiv \left| \vec{x_0} \right|$, so we have the geometry of
Figure~\ref{f:circle}.

By dropping a perpendicular from segment OB to point A (shown as
line segment AC), one readily
sees that $s_i \cos{\psi_i} = R - a \cos{\theta_i}$, with $\theta_i$
being the angle between the event and $i^{th}$ PMT
seen from the origin.  Hence the likelihood function becomes
\begin{equation}
\mathcal{L}(\vec{x_0}, t_0; \{(\vec{x_i}, t_i)\}) = \prod_{i=1}^N 
	p\left(t_i - t_0 - \difrac{s_i n}{c}\right) \,
	\difrac{R - a \cos{\theta_i}}{s_i^3}
	\label{e:sph-likelihood}
\end{equation}
where $s_i$ is given by the Law of Cosines,
\begin{equation}
s_i^2 = R^2 + a^2 - 2 a R \cos{\theta_i}.
\label{e:loc}
\end{equation}

\section{Properties of the Likelihood Function at the Origin}

It may be of interest to examine properties of the likelihood function in the particular case of a hypothetical event occurring at the center of a spherical detector.  This allows the general nature of the problem of reconstruction to be understood analytically.  For simplicity, let's assume that the distribution of the time emission of the photons is a Gaussian curve with width equal to the characteristic dispersion time of the scintillator: 
\begin{equation}
p(\tau_e)=\difrac{e^{-\tau_e^2/2\sigma^2}}{\sqrt{2\pi\sigma^2}}; \;
\log p(\tau_e) = \mbox{\rm const} - \difrac{\tau_e^2}{2 \sigma^2}.
\label{e:gaussian}
\end{equation}

The same equation can also be used for the case when the original scintillation light is absorbed and then re-emitted by scintillation fluors in the immediate proximity of the energy deposition point \cite{ctf-light}.  In this case, the dispersion characteristic of the scintillator is effectively broadened by the absorption and re-emission process.

\subsection{Taylor Expansion of the Likelihood Function}
\label{ss:taylor-expansion}

For a point in the detector at a distance $a$ from the center,
in the direction of a particular unit vector $\hat{u}$, the log
likelihood function is
\begin{equation}
\log{\mathcal{L}(a\hat{u},t_0)}= \mbox{\rm const} -\difrac{1}{2\sigma^2} 
  \sum_{i=1}^{N}\left(t_i-t_0-\difrac{s_i n}{c}\right)^2
  + \sum_{i=1}^{N} \log{\difrac{R - a \cos{\theta_i}}{s_i^3}}
\label{e:spec-likelihood}
\end{equation}
where $s_i$ and $\theta_i$ for each PMT are as shown in figure~\ref{f:circle}.
We assume that the number of hit PMTs $N$ is sufficiently large that we can,
with little error, replace this expression by spatial and temporal averages
over the expected angular and time distributions of the PMT hits.  That is
(discarding the constant term),
\begin{equation}
\log{\mathcal{L}(a\hat{u},t_0)} \; \approx \; -\difrac{N}{2\sigma^2} 
  \left< \left(t-t_0-\difrac{s n}{c}\right)^2 \right>
  \; + \; N \left< \log{\difrac{R - a \cos{\theta}}{s^3}} \right>,
\label{e:avg-likelihood}
\end{equation}
where $t$, $s$, $\theta$ are now continuous random variables with the
expected distributions.
We now calculate these averages for a point-like event located in the center
$\vec{x}_0=\vec{0}$ of the detector, occurring at time $t_0 = 0$.

First consider the time average.  The time of flight of photons from the
center to each PMT (assuming minimal scattering) is $Rn/c$, where $n$ is the
index of refraction and $c$ is the velocity of light in vacuum.
This means that the distribution curve of $t$ is $p(t-Rn/c)$.  From the
properties of a Gaussian distribution, the time averages of time-dependent
quantities are
\begin{eqnarray}
\left< t \right>   & = & \difrac{Rn}{c} \\
\left< t^2 \right> & = & \left<t\right>^2 + \sigma_t^2
	= \difrac{R^2n^2}{c^2} + \sigma^2.
\end{eqnarray}

Likewise, since all PMTs are equidistant from an event at the center of a
spherical detector, the distribution of PMT hits should be uniform over the
solid angle.  Hence the spatial averages over quantities dependent upon the
event-to-PMT angle $\theta$ can be found using Equation~(\ref{e:loc}) and
taking the surface integral over the sphere of PMTs:
\begin{eqnarray}
\left< s \right> & = & \difrac{1}{4\pi} \int \mathrm{d}\phi\,
	\mathrm{d}\left(\cos \theta\right)
	\sqrt{R^2 + a^2 - 2aR \cos{\theta}}
	\; = \; R + \difrac{a^2}{3R} \\
\left< s^2 \right> & = & \difrac{1}{4\pi} \int \mathrm{d}\phi\,
	\mathrm{d}\left(\cos \theta\right)
	\left(R^2 + a^2 - 2aR\cos{\theta}\right)
        \; = \; R^2+a^2
\end{eqnarray}

Finally, we observe that for a point-like event in the center of a uniform
sphere of PMTs, there is no correlation between the expected spatial
distribution of $s$ and temporal distribution of $t$; that is,
$\left< s t \right> = \left< s \right> \left< t \right>$.  This and
the above equations allow us to evaluate
\begin{eqnarray}
\left<\left(t-t_0-\frac{s n}{c}\right)^2\right>
        & = & \left<t^2 + t_0^2 + \frac{s^2n^2}{c^2}
        - 2t t_0 - 2t\frac{s n}{c} + 2t_0\frac{s n}{c}\right> \nonumber \\        & = & \frac{R^2 n^2}{c^2} + t_0^2 + (R^2 + a^2)\frac{n^2}{c^2} - 2\frac{Rn}{c} t_0 \nonumber \\
        & & \;\;\;\; -\, 2\frac{Rn^2}{c^2} (R + \frac{a^2}{3R})
        + 2t_0 (R + \frac{a^2}{3R}) \frac{n}{c} \nonumber \\
        & = & \, \mbox{\rm const} + t_0^2 + \frac{n^2}{3c^2}a^2
        + \frac{2n}{3cR}a^2 t_0
\end{eqnarray}

where the constant term contains whatever does not depend explicitly on $t_0$ and $a$.

The quantity averaged over in the last term of Equation~(\ref{e:avg-likelihood}), again substituting in
Equation~(\ref{e:loc}), becomes
\begin{eqnarray}
\log{\difrac{R - a \cos \theta_i}{s_i^3}} & = &
\log{\left(\difrac{R - a\cos\theta_i}
	{\left( R^2 + a^2 - 2aR \cos\theta_i \right)^{3/2}}\right)}\nonumber \\
&=& -2\log{R} + \difrac{2a}{R}\cos\theta_i
		+ \difrac{a^2}{2R^2} \left( 5 \cos^2 \theta_i - 3 \right)
		+ ...
\end{eqnarray}
with the last equality above being the expansion into a Taylor series in $a/R$.

By once again averaging the expected distributions in $s$ and $\theta$ 
over the solid angle, the result, obtained to second order in $a/R$, is
determined to be
\begin{equation}
\left< \log{\difrac{R - a \cos \theta}{s^3}} \right> \approx
	\mbox{\rm const} - \difrac{2a^2}{3R^2}.
\end{equation}
The complete likelihood function for an event at the center of a spherical detector, to second order in $a/R$, is thus
\begin{equation}
\log{\mathcal{L}}(a\hat{u}, t_0) \approx
	\mbox{\rm const} - N \left[ \difrac{1}{2 \sigma^2}
	\left(t_0^2 + \difrac{n^2}{3c^2} a^2 + \difrac{2n}{3cR}a^2 t_0\right)
	+ \difrac{2}{3R^2} a^2 \right].
	\label{e:likelihood-at-center}
\end{equation}

\subsection{Likelihood Function Maximum and Resolutions}

Solving for the maximum of the likelihood function and requiring $|a| < R$ gives the expected solutions:
\begin{equation}
\left\{ 
\begin{array}{l}
\difrac{\partial}{\partial t_0}\log \mathcal{L} = 0 \\
\difrac{\partial}{\partial a}\log \mathcal{L} = 0
\end{array}
\right.
\Longleftrightarrow
\left\{
\begin{array}{l}
t_0 = 0 \\
a = 0
\end{array}
\right.
\end{equation}

We next ask about the expected resolution of the detector.
Notice that the information matrix is diagonal because the off-diagonal
terms, $-\partial^2 (\log{\mathcal{L}}) / \partial a \partial t_0$, are
zero when $a = t_0 = 0$.  The theoretical resolutions of the detector in
space and time are therefore given by reciprocals of the second derivatives
of the likelihood function:
\begin{equation}
\left\{ 
\begin{array}{l}
\delta t_0 = \left( -\difrac{\partial^2 \log{\mathcal{L}}}{\partial t_0^2}
		\right)^{-1/2} = \difrac{\sigma}{\sqrt{N}} \\
\delta a = \left( -\difrac{\partial^2 \log{\mathcal{L}}}{\partial a^2}
		\right)^{-1/2} =
	\left( \difrac{Nn^2}{3c^2 \sigma^2} \, + \,
		\difrac{4N}{3R^2} \right)^{-1/2}
\end{array}
\right.
\label{e:resolution}
\end{equation}
When the detector dimensions are much larger than the scintillator dispersion time, $R \gg c \sigma / n$, we can approximate $\delta a \approx \sqrt{\difrac{3}{N}} \difrac{c\sigma}{n}$.  (It should be noted that this does not take into
account scattering effects, which become increasingly important with
larger detectors.)

Because of the spherical symmetry of the problem,
$\delta a$ can be used as a stand-in for any of the three Cartesian spatial
resolutions $\delta x_0$, $\delta y_0$, $\delta z_0$.  One may, for instance,
make the substitution $a^2 = x_0^2 + y_0^2 + z_0^2$ in
Equation~(\ref{e:likelihood-at-center}) and obtain the same results for
the resolution in each Cartesian coordinate.

\subsection{Pattern Matching}

In case of use of a liquified noble gas as scintillator, as in the new generation of solar neutrino detectors \cite{clean,xmass}, Rayleigh scattering of the ultraviolet scintillation photons plays an important role.  The photons are scattered intensely by the medium, such that they effectively diffuse out of the medium with a very long dispersion time; then $R \gg c \sigma / n$ is no longer valid.  In this case, the information carried by the time of flight method about the original position of the events becomes less reliable.  However, it is still possible to reconstruct the original position of the event by taking into account that the density of hits on the PMTs decreases with the inverse of the squared distance from the point where the energy is deposited~\cite{clean-rec}.

Suppose that we have no timing information, so our only information about an event is the pattern of hit PMTs.  In this case, the likelihood function simply determines the position of the event. It does not depend  on time and cannot be used to reconstruct the time itself.  We may set the function $p(\tau_e)$ to be constant and ignore it:
\begin{equation}
\log{\mathcal{L}(a\hat{u})}= \mbox{\rm const}
  + \sum_{i=1}^{N} \log{\difrac{R - a \cos{\theta_i}}{s_i^3}}.
\end{equation}

By the same methods as above, we obtain
\begin{equation}
\log{\mathcal{L}}(a\hat{u}) \approx \mbox{\rm const} - \difrac{2N}{3R^2} a^2
\end{equation}
for the second-order Taylor expansion in $a/R$ of the likelihood function for an event at the detector center.  In this case we find
\begin{equation}
\difrac{\partial}{\partial a}\log \mathcal{L} = 0
\Longleftrightarrow
a = 0,
\end{equation}
and for the resolution,
\begin{equation}
\delta a = \left( -\difrac{\partial^2 \log{\mathcal{L}}}{\partial a^2}
		\right)^{-1/2} =
          \sqrt{\difrac{3}{N}} \difrac{R}{2}.
\label{e:pattern-resolution}
\end{equation}

Recall Equations~(\ref{e:likelihood-at-center}) and~(\ref{e:resolution}) in
the case where timing information {\it is} available:
\begin{eqnarray*}
\log{\mathcal{L}}(a\hat{u}, t_0) & \; \approx \; &
	\mbox{\rm const} - \difrac{2N}{3R^2}a^2
	- \difrac{N}{\sigma^2}
	\left(t_0^2 + \difrac{n^2}{3c^2} a^2 + \difrac{2n}{3cR}a^2 t_0\right) \\
\delta a & \; = \; &
	\left( \difrac{Nn^2}{3 c^2 \sigma^2} \, + \,
	\difrac{4N}{3R^2} \right)^{-1/2}
	\approx \; \sqrt{\difrac{3}{N}} \difrac{c\sigma}{n}.
\end{eqnarray*}
We see that use of timing information improves spatial resolution significantly when the scintillator dispersion time is much less than the travel time for light to cross the detector.
In a liquid noble gas detector, the scintillator time dispersion is very
broad due to the amount of internal Rayleigh scattering of scintillation light.
Nevertheless, use of even the small amount of timing information available has
been shown to improve the spatial resolution by a large
fraction~\cite{neon-darkmatter}.

\subsection{Comparison to Observed Resolutions}

\begin{table}[t!]
\begin{center}
\begin{tabular}{lcrcrcrrr} \hline \hline
Detector & $R$ & $T$ & $n$ & $\sigma$
& $\epsilon$ & $N$ & Pred. & Obs. \\
& [m] & & & [ns] &
[pe] & & \multicolumn{2}{c}{$\delta a$\,[cm]} \\ \hline
\multicolumn{8}{l}{Organic scintillator detectors} \\
CTF, $^{214}$Po $\alpha$~\cite{ctf,ctf-results} & 3.3 & 100 & {\it 1.8} &
{\it 5.1} & 225 & 90 & 12.0 & 12.3 \\
Borexino, 1\,MeV $e^-$ MC~\cite{bx} & 6.5 & 2240 & 1.5 &
{\it 5.1} & 400 & 366 & 8.8 & 8.0 \\ \hline
\multicolumn{8}{l}{Hypothetical
	$\ell$Ne detector, 100\,keV $e^-$ MC~\cite{neon-darkmatter}} \\
Spatial data only & 3.0 & 1832 & - & - &
{\it 243} & {\it 243} & 16.7 & 17.0 \\
Timing included & " & " & 1.2 &
{\it 10} & {\it 162} & {\it 155} & 15.0 & 13.6 \\
\hline
\end{tabular}
\end{center} 
\caption{Comparison of the predicted resolutions of three liquid
scintillator detectors with
the values determined experimentally or by Monte Carlo (MC) methods.
See the text for meanings of the columns and comments on values in
{\it italics}.}
\label{t:resolutions}
\end{table}

Experimentally, the position resolution of a detector can be determined in
several ways.  The simplest and most common is the use of a calibration source.
In cases when the detector has not yet been built, Monte Carlo methods are of
course the only method that can be used.  The detector resolutions obtained
from experimental results for CTF, and Monte Carlo tests of Borexino
and a hypothetical liquid neon dark matter detector~\cite{neon-darkmatter},
are shown
in the last column of Table~\ref{t:resolutions}.  For comparison, the physical
attributes of the detectors and the predicted resolutions $\delta a$ from
Equation~(\ref{e:resolution}) are shown in the other columns of the table.  As
above, $R$ is the detector radius, $T$ the total number of PMTs, $n$
the scintillator index of refraction, and $\sigma$ the scintillator dispersion
time.  The average number of photoelectrons detected in each event from the
source used is denoted by $\epsilon$.

$N$ is determined in most cases as follows.  In detectors using a
time-of-flight position reconstruction method, each PMT can
measure the arrival time only of the first photon to strike it.  This
difficulty will be discussed more thoroughly in Section~\ref{s:orderstat}.  The
immediate consequence is that $N$ is a measure of the number of hit PMTs rather
than the total number of detected photoelectrons.  Basic probability tells us
that given an event in which $\epsilon$ photoelectrons are detected, the
expected number of hit PMTs is
\begin{equation}
\left< N \right> = T \left[1 - \left(\difrac{T - 1}{T}\right)^\epsilon \right].
\label{e:N-from-epsilon}
\end{equation}
Note, however, that for the spatial hit pattern, every photoelectron
contributes to our knowledge, even for multiple hits on a single PMT.  This
implies that the term $4N/3R^2$ in the expression for $\delta a$ in
Equation~(\ref{e:resolution}) should in fact include $\epsilon$, not $N$.
In calculating the predicted values of $\delta a$ in Table~\ref{t:resolutions},
we therefore use the modified expression
\begin{equation}
\delta a = \left( \difrac{Nn^2}{3c^2 \sigma^2} \, + \,
		\difrac{4\epsilon}{3R^2} \right)^{-1/2}.
\label{e:resolution-modified}
\end{equation}

Some comments on idiosyncracies of the individual detectors are in order.  The
value of $n$ of 1.8 tabulated for the CTF is an ``effective index of
refraction.'' In
fact, the CTF volume is partly water ($n = 1.33$) and partly organic
scintillator ($n = 1.5$); this ``effective index'' is an attempt to account
for refraction at the interface between the two fluids.  Refraction
causes light to travel a greater distance from event to PMT than it would
through a single medium, so the ``effective $n$'' is higher than that of
either pure fluid.  Additionally, note that the observed value
of $\delta a$ for the CTF takes into account only the spread in $x$ and $y$
coordinates; the CTF source had the shape of a cylinder, extended in~$z$.

In the hypothetical liquid neon detector described in
reference~\cite{neon-darkmatter}, events have a prompt component
(relative intensity 2.0) and a delayed component (relative intensity 1.0)
of scintillation light.  For the Monte Carlo simulation taking into account
only the spatial pattern of PMT hits (``spatial data only'' row of
Table~\ref{t:resolutions}), both components contribute useful data.
In that case the photoelectron yield is 2428\,pe/MeV, 1.5 times the prompt
light yield of 1619\,pe/MeV (10791.7\,photons/MeV $\times$ 20\% quantum
efficiency $\times$ 75\% geometric coverage) quoted in the reference.  For the
position reconstruction calculated from the spatial pattern only, we use
$N = \epsilon_{total} \equiv \epsilon_{prompt} + \epsilon_{delayed}$ in
Equation~(\ref{e:pattern-resolution}).

Calculation of the expected resolution in the liquid Ne detector is trickier
when timing information is included (``timing included'' row of
table~\ref{t:resolutions}).  The two terms contributing to $\delta a$ in
Equation~(\ref{e:resolution-modified}) must be evaluated with different values
for $\epsilon$.  The term $4\epsilon/3R^2$ comes from the spatial hit pattern
and so uses $\epsilon_{total} = 243$, while the timing-dependent term
$Nn^2/3c^2\sigma^2$ includes only the prompt component of scintillation light,
and thus uses $\epsilon_{prompt} = 162$, with $N = 155$ derived from
Equation~(\ref{e:N-from-epsilon}).

The source of the largest potential errors in the predictions of
Table~\ref{t:resolutions} is the value of the scintillator
dispersion,~$\sigma$.  The true scintillator dispersion function of a detector
$p(\tau_e)$ is not actually a Gaussian, so the use of
Equation~(\ref{e:resolution}) is only an approximation.  The value of 5.1~ns
used for $\sigma_{CTF}$ is obtained from the fit to CTF data described in
reference~\cite{ctf-light} with the parameters shown in Figure~6 of that paper,
sampled at 1~ns intervals and fit to a Gaussian only.  (The same scintillator
dispersion function was used in the Borexino Monte Carlo simulations.)
Nevertheless, the predicted, observed and Monte Carlo values of the position
resolution are in quite good agreement.  For the liquid Ne detector,
$\sigma$ was estimated at 10\,ns, based on Figure~7 of
reference~\cite{neon-darkmatter}, as 1/2 the difference between times with
probability values equal to $e^{-0.5}$ times the value at the peak.  One
could plausibly estimate this value of $\sigma$ to be anywhere in the range
5.5 to 15\,ns, yielding estimates of $\delta a$ from 12.6 to 15.9\,cm.
This range brackets the Monte Carlo simulation nicely.

\section{Multiple PMT Occupancy and Order Statistics}
\label{s:orderstat}

So far it has largely been assumed that the occupancy of each PMT in the
detector is at most one.  If the detector has the capability to measure the
time at which {\it every} photon hits a given PMT, or if the detector
(as with some of the proposed noble gas detectors) has no timing capability
at all, then the assumption may be lifted
with no effect, except that some of the $\vec{x_i}$ (and hence $\theta_i$ and
$s_i$) will be identical in Equation~(\ref{e:sph-likelihood}).  For a
detector with timing capabilities, however, it is more
likely that the detector only has the capability to measure the
arrival time of the {\it first} photon to reach each PMT.  The probability
function of the first photon to reach a PMT is not the same as that of a random
photon reaching the same PMT; it
is biased toward earlier times.  To account for this bias, the scintillator
response function $p(\tau_e)$ must be corrected.

\subsection{Correcting for Timing Bias}

Let the probability function of the first photon to reach a PMT,
out of the $n$ photons reaching that PMT from an event, be represented by
$p_n(\tau_e)$.
This is known as the ``first order statistic.''
Naturally, $p_1(\tau_e) \equiv p(\tau_e)$.
In general, the corrected scintillator response function $p_{corr}$ would then
be some linear combination of the first order statistics,
\begin{equation}
p_{corr}(\tau_e) =
        \sum_{n=1}^\infty
        p_n(\tau_e) \times
        \mathrm{P}(n\, \mathrm{photons\, hit\, the\, PMT}),
\end{equation}
and an {\it a priori} guess would have to be made for the probability that
each possible number of photons had hit the PMT.  For simplicity, let us assume
that the number of photons striking each PMT for an event is known (in
Borexino, for instance, this is determined via ADC channels separate from
the timing channels).  We can then set $p_{corr}$ equal to the function
$p_n(\tau_e)$.

\begin{figure}[!t]
\begin{center}
 \psfrag{tau}{$\tau_e / \sigma$}
 \psfrag{poftau}{$p(\tau_e)$}
 \psfrag{pnoftau}{$p_n(\tau_e)$}
 \psfrag{n2}{$n=2$}
 \psfrag{n3}{$n=3$}
 \psfrag{n5}{$n=5$}
 \psfrag{n10}{$n=10$}
 \epsfig{file=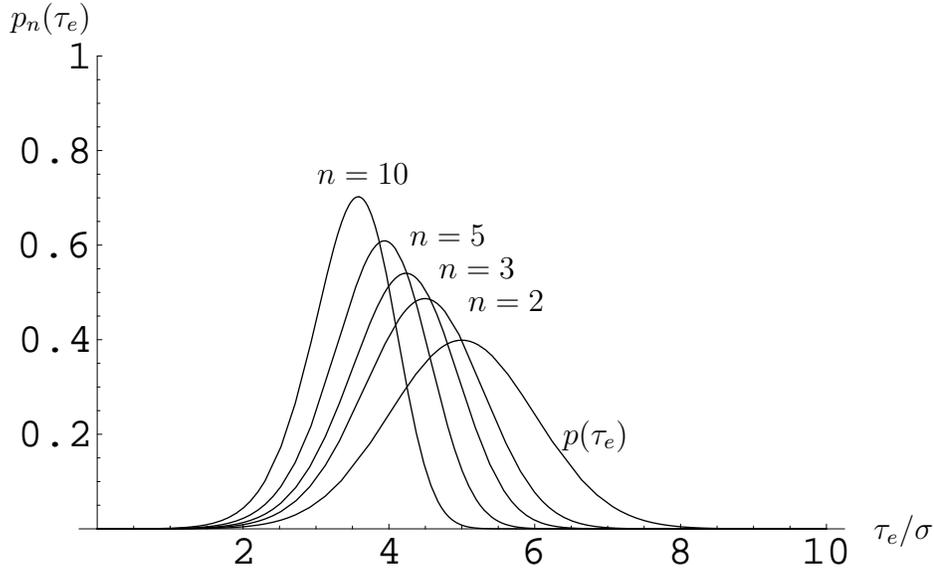,width=5in}
\end{center}
\caption{
A hypothetical Gaussian scintillator response function $p(\tau_e)$,
and its first order statistics for increasing values of $n = 2, 3, 5, 10$. Note
how as $n$ increases, the corrected response function narrows and shifts toward
earlier times.  The time axis is shown in units of the scintillator
dispersion time $\sigma$.}
\label{f:order-statistics}
\end{figure}

It remains only to calculate $p_n(\tau_e)$ given $p(\tau_e)$ and $n$.
Number the emission time of the $n$ photons detected by a given PMT in some
specific but randomly chosen order (for instance, in order of increasing
longitude of their emission directions), $\tau_1, \ldots, \tau_n$.
Also number them in order of
increasing emission time, $s_1, \ldots, s_n$.  Then $p_n(\tau_e)$ is
the probability function of the randomly chosen emission time
$\tau_1$ given that $s_1 = \tau_1$:
\begin{eqnarray}
p_n(\tau_e)\, \mathrm{d}\tau_e \;&=&\; \mathrm{P}(\tau_1 \in
[\tau, \tau + \mathrm{d}\tau] | \tau_1 = s_1) \nonumber \\
&=&\; \mathrm{P}(\tau_1 = s_1 | \tau_1 \in [\tau, \tau + \mathrm{d}\tau])
\times \frac{\mathrm{P}(\tau_1 \in [\tau, \tau + \mathrm{d}\tau])}
{\mathrm{P}(\tau_1 = s_1)} \nonumber \\
&=&\; \frac{p(\tau_e)\, \mathrm{d}\tau_e}{(1/n)}\,
\mathrm{P}(\tau_1 = s_1 | \tau_1 \in [\tau, \tau + \mathrm{d}\tau]),
\end{eqnarray}
where the second equality is once again due to Bayes' Theorem.  The
probability in the last line above is just the probability that every
other photon has a later arrival time than the randomly selected value $\tau_1$:
\begin{eqnarray}
\mathrm{P}(\tau_1 = s_1 | \tau_1 \in [\tau, \tau + \mathrm{d}\tau])
\;&=&\; \prod_{i = 2}^n \mathrm{P}(\tau_i > \tau_1 | \tau_1 \in
[\tau, \tau + \mathrm{d}\tau]) \nonumber \\
&=&\; \mathrm{P}(\tau_2 > \tau_1 | \tau_1 \in
[\tau, \tau + \mathrm{d}\tau])^{n-1} \nonumber \\
&=&\; \left[ \int_{\tau_e}^\infty p(\tau_e')\, \mathrm{d}\tau_e' \right]^{n-1}.
\end{eqnarray}

Hence (letting $F(\tau_e) \equiv \int_{-\infty}^{\tau_e} p(\tau_e)\,
\mathrm{d}\tau_e$ represent the cumulative distribution function of
$\tau_e$), the first order statistic of $p(\tau_e)$, if $n$ photons are
detected by a given PMT, is
\begin{equation}
p_n(\tau_e) \; = \; n p(\tau_e)\, \left[1 - F(\tau_e) \right]^{n-1}.
\end{equation}

Graphs of the first order statistics of a representative scintillator
response function are shown in Figure~\ref{f:order-statistics} for values of
$n$ equal to 1, 2, 3, 5, and 10.  (The specific response function shown is a
Gaussian, Equation~(\ref{e:gaussian}) offset by five units of $\sigma$
from time zero.)  Note how as
$n$ increases, the time distribution of the first PMT hit narrows and shifts
toward earlier times.

\subsection{Effects on Detector Resolution}

One may ask about the effect of this correction on the likelihood function
and spatial resolution.  Consider again the case of a Gaussian scintillator
time response function.  We have
\begin{equation}
\log p_n(\tau_e) \; = \;
	\mbox{\rm const} + \log p(\tau_e) + (n - 1)\log [1 - F(\tau_e)].
\end{equation}
Substituting in $F(\tau_e) = (1 + {\rm erf} (\tau_e / \sigma \sqrt{2}))/2$, the
Taylor expansion to second order in $\tau_e$ becomes
\begin{equation}
\log p_n(\tau_e) \; = \; \mbox{\rm const} -
	(n - 1)\sqrt{\frac{2}{\pi}}\frac{\tau_e}{\sigma}
	- \left(\frac{1}{2} + \frac{n-1}{\pi}\right)\frac{\tau_e^2}{\sigma^2}
	+ O(\tau_e^3).
\end{equation}
That is, the first photon detected at each PMT contributes to the log of the
likelihood function in the amount of $-\tau_e^2/2\sigma^2$, but each additional
photon contributes only in the amount of $-\tau_e^2/\pi\sigma^2$ (plus a term
linear in $\tau_e$ which has relatively little effect on the resolution
for a large detector); compare to Equation~(\ref{e:gaussian}).  The
resolution is better than if the corrected scintillator response function were
not used, but still poorer than if the time of arrival of every photon could be
measured.

Suppose that the total number of photons detected is $\epsilon$, by $N$ PMTs,
and in particular that the $i^{th}$ PMT sees $n_i$ photons.  Denoting the
emission time by $\tau_e^i \equiv t_i-t_0-s_i n / c$, the general
likelihood function is then
\begin{eqnarray}
\log{\mathcal{L}(a\hat{u},t_0)} &\;=\;& \mbox{\rm const} 
\,-\, \difrac{1}{\sigma^2} 
  \sum_{i=1}^{N} \left( \difrac{1}{2} + \difrac{n_i - 1}{\pi} \right)
  		 \left(\tau_e^i\right)^2 \nonumber \\
&\;& -\, \difrac{1}{\sigma}\sqrt{\difrac{2}{\pi}} \sum_{i=1}^N
	(n_i - 1)\, \tau_e^i
\,+\, \sum_{j=1}^{\epsilon} \log{\difrac{R - a \cos{\theta_j}}{s_j^3}}.
\end{eqnarray}
Define the excess photon multiplicity as $\delta \equiv (\epsilon - N)/N$.  The
likelihood function in the limit of homogeneous PMT coverage as
$N \rightarrow \infty$, for an event at the detector center, becomes
\begin{eqnarray*}
\log{\mathcal{L}(a\hat{u},t_0)} &\;=\;& \mbox{\rm const} 
\,-\, \difrac{N}{\sigma^2} \left( \difrac{1}{2} + \difrac{\delta}{\pi} \right)
	\left< \left( \tau_e^i \right)^2 \right>\\
&\;& -\, \difrac{N\delta}{\sigma} \sqrt{\difrac{2}{\pi}} \left< \tau_e^i \right>
\,+\, N (\delta + 1)\, \left< \log \difrac{R - a \cos{\theta_j}}{s_j^3}
	\right>.
\end{eqnarray*}
Running through calculations analogous to those of Section~\ref{ss:taylor-expansion},we finally obtain the explicit function
\begin{eqnarray}
\log{\mathcal{L}(a\hat{u},t_0)} &\;=\;& \mbox{\rm const}
\,-\, \difrac{N}{\sigma^2} \left( \difrac{1}{2}
	+ \difrac{\delta}{\pi} \right)\left( t_0^2 + \difrac{n^2}{3c^2} a^2
	+ \difrac{2n}{3cR} a^2 t_0 \right)  \nonumber\\
&\;&    -\, \difrac{N \delta}{\sigma} \sqrt{\difrac{2}{\pi}}
		\left(t_0 + \difrac{n}{3cR} a^2 \right)
	\,-\, N (\delta + 1)\difrac{2}{3R^2} a^2 .
\end{eqnarray}

In the limit $c \sigma / R \rightarrow 0$ (that is, for a very large detector
compared to the width of the scintillator response function), it can be shown
that the spatial resolution at the center of a detector, with $N$ and $\delta$
varying while holding $\epsilon$ constant, is proportional to
$\sqrt{\pi(1+\delta)} / \sqrt{\pi+2\delta}$.  Hence the resolution of an event
with an average photon multiplicity of $\delta = 0.5$~excess photons per PMT is
6.7\% worse than if PMTs could detect the arrival time of every photon.  With
$\delta = 1$~excess photon per PMT (every hit PMT seeing an average of 2
photons), the resolution is 10.5\% worse.  In the limit of large $\delta$
(for instance with a high-energy event), the
resolution reaches an asymptote of $\sqrt{\pi/2}$ times (about 25.3\% worse)
that of an ideal detector observing an event of equal energy.

Realistically, construction of an ideal detector, one that measures the time
of arrival for every photon, would be non-trivial.  One may on the other hand
ask, given a detector capable of measuring time of
arrival only for the first photon at each PMT, how the use of the statistically
corrected scintillator dispersion function improves the results over the use of
an uncorrected function.  This comparison is equivalent to fixing $N$ while
(for the uncorrected dispersion function) setting $\delta$ to zero.  In this
case, the use of the corrected dispersion function is an improvement by the
factor $\sqrt{\pi} / \sqrt{\pi + 2\delta}$ (recall that smaller resolutions are
better).  For $\delta = 0.5$, the reciprocal of the improvement factor is
1.15, and for
$\delta = 1$, it is 1.28; for large $\delta$, it would theoretically
improve without bound.  This analysis even leaves aside the fact that for
events offset from the center of the detector, use of the uncorrected
scintillator dispersion function will produce a statistically biased position
estimate.

\section{Conclusions}

We analyzed the resolution of spherical, optical, non-imaging scintillation based detectors in reconstructing the position of point-like events, limiting the analytic derivation to the case of events near the center of the detector.  We found that the fundamental length scale of the resolution given by the time of flight method is proportional to the product of the speed of light in the medium and the dispersion time at the scintillation emission, as in $\delta a \approx \sqrt{\difrac{3}{N}} \difrac{c\sigma}{n}$.

In case the dispersion of the scintillation photons arrival times grows above the ratio of the speed of light to the detector radius, the time of flight method no longer gives relevant information about the point of origin of the event.  The position of the event can still be determined by the analysis of the density of hits, and in this case the fundamental resolution is set by the radius of the detector, as in $\delta a = \sqrt{\difrac{3}{N}} \difrac{R}{2}$.

Finally, we made some comments on the need to correct the scintillation
dispersion function in the common case where PMT hit timing information is only
available for the first photon to strike each PMT.  In this case, even with a
corrected scintillation dispersion function, the spatial resolution will
be up to 25~percent worse for high-energy events compared to a
similar detector capable of measuring timing information for all photons.

\section{Acknowledgments}

The authors are grateful for the many helpful suggestions and comments of
Kevin Coakley, Dan McKinsey, and Andrea Pocar.

\end{document}